\pdfoutput=1 

\documentclass[%
 superscriptaddress,
preprint,
 amsmath,amssymb,
 aps,
 pra,
]{revtex4-1}

\usepackage{graphicx}               
\usepackage{dcolumn}                
\usepackage{bm}                     
\usepackage{siunitx}
\usepackage{tabularx}
\usepackage{dsfont}					
\usepackage{mathrsfs}				
\usepackage{float}
\usepackage{hyperref}               

\newcommand{\comment}[1]{}			

\newcommand{\be}[0]{\begin{equation}}	
\newcommand{\ee}[0]{\end{equation}}

\begin{document}


\title{Evanescent Modes and Step-like Acoustic Black Holes}
\author{Jonathan Curtis}
\affiliation{Joint Quantum Institute, University of Maryland, College Park, MD 20742, USA}
\affiliation{Condensed Matter Theory Center, University of Maryland, College Park, MD 20742, USA}
\author{Gil Refael}
\affiliation{Institute of Quantum Information and Matter, California Institute of Technology, Pasadena, CA 91125, USA}
\affiliation{Department of Physics, California Institute of Technology, Pasadena, CA 91125, USA}
\author{Victor Galitski}
\affiliation{Joint Quantum Institute, University of Maryland, College Park, MD 20742, USA}
\affiliation{Condensed Matter Theory Center, University of Maryland, College Park, MD 20742, USA}

\date{\today}

\begin{abstract}
We consider a model of an acoustic black hole formed by a quasi-one dimensional Bose-Einstein condensate with a step-like horizon.
This system is analyzed by solving the corresponding Bogoliubov-de Gennes equation with an appropriate matching condition at the jump.
When the step is between a subsonic and supersonic flow, a sonic horizon develops and in addition to the scattering coefficients we compute the distribution of the accompanying analogue Hawking radiation.
Additionally, in response to the abrupt variation in flow and non-linear Bogoliubov dispersion relation, evanescent solutions of the Bogoliubov-de Gennes equation also appear and decay out from the horizon.
We bound this decay length and show that these modes produce a modulation of observables outside the event horizon by their interference with outgoing Hawking flux. 
We go further and find specific superpositions of ingoing eigenmodes which exhibit coherent cancellation of the Hawking flux outside the horizon but nevertheless have evanescent support outside the black hole. 
We conclude by speculating that when quasiparticle interactions are included, evanescent modes may yield a leakage of information across the event horizon via interactions between the real outgoing Hawking flux and the virtual evanescent modes, and that we may expect this as a generic feature of models which break Lorentz invariance at the UV (Planck) scale. 
\end{abstract}

\maketitle

\section{\label{sec:intro}Introduction}
The existence of black holes was one of the first surprising and novel predictions to emerge from Einstein's geometric theory of gravity.
Though initially their existence was a point of contention, it is now well established that black holes exist and play a formative role in the large-scale dynamics of the universe~\cite{Bambi18}.
The interplay between quantum mechanics and these exotic spacetime solutions has uncovered a number of important open problems, with fundamental consequences for theories of quantum gravity an cosmology~\cite{Giddings17}.
Perhaps nowhere is this more evident than in Stephen Hawking's 1974 semiclassical calculation which predicted that black holes constantly emit a flux of thermal quanta~\cite{Hawking74,Hawking75}.
Consequences of this thermal radiation have since raised a number of fundamental questions regarding the interplay of quantum mechanics and gravity, including what is known as the ``black hole information paradox~\cite{Susskind06,Sully13}."

Though it is grounded in widely accepted physical principles, observation of astrophysical Hawking radiation seems to be impossible in the near-future at least. 
In 1981, W.G. Unruh proposed that in lieu of observation of Hawking radiation by an astrophysical black hole, the process of Hawking radiation and black-hole evaporation could be effectively simulated in a laboratory~\cite{Unruh81}. 
This proposal, expanded upon extensively by G.E. Volovik as pertaining to superfluid Helium~\cite{Volovik09}, relies on the observation that at long wavelengths sound waves propagating through a fluid are described by the same equations of motion as a scalar boson propagating through a curved spacetime~\cite{Stone00,Visser01,Visser11,Galitski18}.
A simple model which illustrates this physics is that of a Bose-Einstein condensate (BEC), where the condensate flow plays the role of the spacetime metric while the quantum fluctuations (e.g. phonons) are mapped onto matter fields residing in this spacetime. 
If the condensate velocity exceeds the local speed of sound this effective spacetime develops an event horizon, forming a sonic black/white hole. 
In exact analogy with Hawking's calculations for an astrophysical black hole, the sonic horizon formed in a condensate should then emit a thermal flux of phonons. 

Since Unruh's initial observation, there have been numerous proposals for testing-by-analogy various predictions of semiclassical gravity and cosmology using table-top scale experiments.
These employ a range of media including liquid Helium~\cite{Volovik98,Volovik09,Volovik13}, trapped Bose-Einstein condensates (BECs) and ultra-cold atoms~\cite{Zoller00,Visser01,Schutzhold04,Fischer04,Fischer06,Fischer17,Naderi18,Demler18,Campbell18}, electromagnetic waveguides~\cite{Unruh05}, spintronic materials~\cite{Rousseaux11,Duine17}, exciton-polariton condensates~\cite{Amo15}, non-linear optical media~\cite{Piwnicki00}, and even water wave-tanks, where signatures of Hawking radiation still manifest themselves through the classical correlation functions~\cite{Rousseaux16}.
Recent experiments by Jeff Steinhauer have purportedly generated and observed signatures of self-amplifying Hawking radiation~\cite{Steinhauer14,Clark17} and its entanglement~\cite{Steinhauer16} in an ultra-cold BEC.
Crucially, these experiments do not attempt to detect the Hawking radiation by directly measuring its temperature (which is typically too small to effectively measure), but instead measure non-local density-density correlations which arise due to the Hawking emission~\cite{Carusotto08,Kravtsov09,Pavloff12}.
These observables, which may be measured in the lab, cannot be measured for real black holes since they involve measuring correlations across the event horizon. 

It is an interesting question to consider how all these disparate theories, none exhibiting true Lorentz invariance, differ in their low-energy effective descriptions.
Often, the absence of Lorentz invariance at the UV scale manifests itself through the quasiparticle dispersion relations which exhibit either superluminal~\cite{MacherParentani09D,Parentani10, Rinaldi11,Sols11} or subluminal propagation at higher momenta~\cite{Jacobson96,Corley97,Jacobson99,Fischer06,Unruh13,Uhlmann08,Naderi18,Leonhardt18}. 
At low energies, all models seem to predict the same thermal occupation function first obtained by Hawking~\cite{Ohberg03B,MacherParentani09A,Carusotto09,Pavloff13}.
Deviations become apparent only at higher energies/momenta, where departures from this thermal occupation can be observed~\cite{Corley97,MacherParentani09A,Carusotto09,Rinaldi11}. 

In this work we will consider a BEC model of analogue gravity, with the goal being to study how the emergent spacetime responds to regions of large effective spacetime curvature. 
A major conclusion of ours is that when the background metric varies over a sufficiently abrupt length scale it becomes possible to observe the emergence of evanescent field modes outside the sonic event horizon, which in turn can effect local observables. 
To motivate this, consider identifying a sonic black hole with an actual astrophysical black hole of equal Hawking temperature.
In a Schwarzschild black hole, the temperature $T_{\textrm{astro}}$ (in units with $\hbar = k_B = 1$) is related to the mass $M$ by Hawking's formula
\[
T_{\textrm{astro}} = \frac{c^3}{8\pi G_N M}
\]
where $c$ is the speed of light and $G_N$ is the Newton gravitational constant. 
For a sonic black hole, we invoke Unruh's result~\cite{Unruh81}, whereby we find that the temperature of the sonic black hole $T_{\textrm{sonic}}$ is related to the fluid velocity gradient by 
\[
T_{\textrm{sonic}} = \frac{1}{2\pi} \bigg|\frac{\partial v}{\partial r}\bigg|_{\textrm{horizon}} 
\]
where the derivative is understood as being taken in the direction normal to the event horizon, at the horizon.
Thus, identifying these two temperatures implies that the mass of the black hole is related to the inverse of the velocity gradient.
If we wish to study the analogue of black holes which are evaporating towards the Planck mass scale, we must understand what happens to the sonic horizon as the velocity gradient increases towards the UV dispersion scale.
In the sonic black hole model, large flow gradients may be modeled most simply by considering a step-like system.
In fact, such configurations have been studied before~\cite{Rinaldi11,Pavloff12,Pavloff13}, though typically the emphasis is placed on obtaining the form of the universal low-energy Hawking distribution function, which is by now well understood. 
In this work, we will instead primarily focus on the near-horizon physics, which has seen comparatively little attention due to its generically non-universal nature. 

Having motivated the step-like model, the remainder of this paper will be structured as follows.
In Section~\ref{sec:formalism}, we will introduce our model and the associated Bogoliubov-de Gennes formalism used to analyze it.
We will then proceed on to consider first the case of a homogeneous flow, presented in Section~\ref{sec:homgeneous}.
In Section~\ref{sec:step} we will set up the step-like system and solve it, extracting both the S-matrix and the actual eigenfunctions of the problem, which contain the evanescent modes.
We will study the properties of these evanescent modes in more detail in Section~\ref{sec:evanescent}, before moving on to the Conclusion in Sec.~\ref{sec:conclusion}, where we highlight some interesting consequences and potential future avenues of research.

\section{\label{sec:formalism}Formalism}
We begin our discussion by considering a model for weakly interacting spinless bosons described by the Hamiltonian 
\begin{equation}
\label{eqn:hamiltonian}
\check{H} = \int d^d r \left( \frac{1}{2m} \nabla\check{\Psi}^\dagger \cdot \nabla \check{\Psi} - \mu \check{\Psi}^\dagger \check{\Psi} + \frac12 g \check{\Psi}^\dagger \check{\Psi}^\dagger \check{\Psi} \check{\Psi} \right), 
\end{equation}
where $d$ is the spatial dimension, $g >0$ is the s-wave interaction constant, and $\mu$ is the chemical potential~\cite{Castin01}.
Here and throughout we use units in which $\hbar = k_B = 1$ and we will distinguish between quantum many-body operators and single-particle differential operators with the use of a check ( $\check{}$ ) and hat ( $\hat{}$ ), respectively. 
The only non-trivial equal-time commutator for the boson field operators $\check{\Psi}(\mathbf{r},t)$ is 
\[
[\check{\Psi}(\mathbf{r},t) , \check{\Psi}^\dagger(\mathbf{r}',t) ] = \delta^{d}(\mathbf{r-r'}).
\]
The resultant many-body dynamics may be described by the Heisenberg equation of motion 
\begin{equation}
    \label{eqn:heisenberg}
    \left( i\partial_t +\frac{1}{2m}\nabla^2 + \mu - g \check{\Psi}^\dagger \check{\Psi} \right) \check{\Psi} = 0. 
\end{equation}
Next, we partition the operator field $\check{\Psi}$ into a classical condensate $\psi = \sqrt{\rho}e^{i\Theta}$ and fluctuations about the condensate via 
\begin{equation}
    \label{eqn:mean-field}
    \check{\Psi}(\mathbf{r},t) = \psi(\mathbf{r},t) \left(\check{\mathds{1}} + \check{\phi}(\mathbf{r},t) \right).
\end{equation}
Note the fluctuations are rescaled by the local condensate so that the equal time commutator for the $\phi$ field reads
\begin{equation}
[\check{\phi}(\mathbf{r},t) , \check{\phi}^\dagger(\mathbf{r}',t) ] = \frac{1}{\rho(\mathbf{r},t)}\delta^{d}(\mathbf{r-r'}).
\end{equation}

Next, we define the superfluid velocity $\mathbf{v} = \frac{1}{m} \nabla \Theta $, in terms of which the mean-field equations of motion become
\begin{equation}
    \label{eqn:Euler-Bernoulli}
    \begin{aligned}
    &   \partial_t \rho + \nabla\cdot(\rho \mathbf{v}) = 0    \\
    &   \mu -\partial_t \Theta -\frac12 m\mathbf{v}^2 - g\rho + \frac{1}{2m\sqrt{\rho} }\nabla^2 \sqrt{\rho} = 0. \\
\end{aligned}
\end{equation}
We insert the ansatz~\eqref{eqn:mean-field} into the Heisenberg equation and apply the mean-field equations of motion~\eqref{eqn:Euler-Bernoulli}. 
To linear order in the operator fields $\check{\phi},\check{\phi}^\dagger$ we get the Bogoliubov-de Gennes (BdG) equation
\begin{equation}
    \label{eqn:fluctuation-EOM}
\left( i\partial_t + i\mathbf{v}\cdot\nabla + \frac{1}{2m\rho}\nabla \cdot \rho \nabla - g\rho \right) \check{\phi} - g\rho \check{\phi}^\dagger  = 0. 
\end{equation}
This is written compactly in terms of the Nambu spinor 
\begin{equation}
	\label{eqn:nambu}
\check{\Phi} = \left(\begin{array}{c}
\check{\phi} 	\\
\check{\phi}^\dagger 	\\
\end{array}\right)
\end{equation}
and the (Nambu-space) Pauli matrices $\tau^1,\tau^2,\tau^3,\tau^0$ as  
\begin{equation}
    \label{eqn:BdG}
    \hat{K}_{\tiny\textrm{BdG}} \check{\Phi} = 0,
\end{equation}
where the BdG kernel is given by the differential operator 
\begin{equation}
    \label{eqn:BdG-kernel}
    \hat{K}_{\tiny\textrm{BdG}} = \left( i\partial_t + i\mathbf{v}\cdot\nabla \right)\tau^3 - \left( g\rho - \frac{1}{2m\rho}\nabla \cdot \rho \nabla \right)\tau^0 - g\rho \tau^1. 
\end{equation}
Additionally, the Nambu spinor may be seen to have the self-conjugate property
\begin{equation}
\label{eqn:self-conj}
\overline{\check{\Phi}}(\mathbf{r},t) \equiv \tau^1\check{\Phi}(\mathbf{r},t)^\dagger =  \check{\Phi}(\mathbf{r},t).
\end{equation}
and at equal times has the commutator
\begin{equation}
	\label{eqn:nambu-commutator}
[\check{\Phi}_\alpha(\mathbf{r},t),\check{\Phi}^\dagger_\beta(\mathbf{r}',t) ] = \frac{1}{\rho(\mathbf{r},t)} \delta^{d}(\mathbf{r}-\mathbf{r}') [\tau^3]_{\alpha\beta} 
\end{equation}
where $\alpha,\beta$ explicitly index the Nambu components. 

We may verify that, provided the background condensate satisfies the continuity equation~\eqref{eqn:Euler-Bernoulli}, the quasiparticle charge
\begin{equation}
    \label{eqn:operator-charge}
    \check{\mathcal{Q}}_{\tiny\textrm{qp}}(t) = \int d^d r \rho (\check{\Phi}^\dagger)^T \tau^3 \check{\Phi}
\end{equation}
and current 
\begin{equation}
    \label{eqn:operator-current}
    \check{\mathbf{J}}_{\tiny\textrm{qp}} = \rho\mathbf{v}(\check{\Phi}^{\dagger})^T \tau^3 \check{\Phi} + \rho\left[-\frac{i}{2m}(\check{\Phi}^{\dagger })^T \nabla\check{\Phi} +\frac{i}{2m} \left(\nabla\check{\Phi}^{\dagger }\right)^T\check{\Phi}\right]
\end{equation}
are conserved under the BdG equations of motion.
Throughout we will carefully distinguish between Hermitian/complex conjugation (which acts element-wise on the spinor components) and Nambu spinor tranposition (which exchanges spinor columns and rows).

We will further restrict our analysis to systems which have time-translational invariance.
In this case, the lab-frame energy $\omega$ is a good quantum number and we the BdG kernel takes the form 
\begin{equation}
    \hat{K}_{\tiny\textrm{BdG}}(\omega) = \tau^3 \left( \omega - \hat{\Omega}_{\tiny\textrm{BdG}} \right),
\end{equation}
which effectively defines the ``BdG Hamiltonian" as the linear differential operator 
\begin{equation}
    \label{eqn:BdG-Hamiltonian}
    \hat{\Omega}_{\tiny\textrm{BdG}} = \tau^3 \left( -\frac{1}{2m\rho} \nabla \cdot \rho \nabla + g\rho \right) + i\tau^2 g\rho - i\mathbf{v}\cdot\nabla .
\end{equation}
In general, the operator $\hat{\Omega}_{\tiny\textrm{BdG}}$ may have complex energy eigenvalues, leading to dynamical instabilities~\cite{Ohberg03A,Castin01,Parentani10,Leonhardt18}.
Though potentially interesting, we will assume that our system does not exhibit these instabilities and that the energy eigenvalues are real.

In order to describe the many-body quantum dynamics of the system, we will first obtain the classical normal modes of the BdG Hamiltonian.
To produce an expansion for the operator $\check{\Phi}$ we will then second-quantize these classical modes. 
Utilizing conservation of the charge defined in Eqn.~\eqref{eqn:operator-charge}, we define a conserved pseudo-inner product~\cite{Jacobson05,MacherParentani09A,MacherParentani09D}
\begin{equation}
    \label{eqn:norm}
    (F,G) = \int d^d r \rho(\mathbf{r}) F^{*T}(\mathbf{r}) \tau^3 G(\mathbf{r}),
\end{equation}
where $F$ and $G$ are two c-number spinor fields.
This product obeys the properties
\begin{equation}
\label{eqn:norm-identities}
    \begin{aligned}
        &   (F,\tau^1 G) = -(\tau^1 F,G)    \\
        &   (F,\tau^2 G) = -(\tau^2 F,G)    \\
        &   (F,\tau^3 G) = +(\tau^3 F,G)    \\
        &   (F,G)^* = (G,F) = (F^*,G^*). \\
    \end{aligned}
\end{equation}
Due to the presence of $\tau^3$ this is not a {\it bona fide} inner product, since we may have $(F,F)<0$ for some modes.

The sign of the norm $(F,F)$, as we will now explain, is closely connected to the creation/annihilation of particles.
To see this, we construct a time-dependent many-body ``wavepacket" operator
\begin{equation}
\label{eqn:mode-operator}
    \check{\mathfrak{a}}_t[F] \equiv \left(F,\check{\Phi}(t) \right)
\end{equation}
from the c-number spinor $F$.
The Hermitian conjugate of this operator may be shown to be
\begin{equation}
\label{eqn:mode-conj}
    \check{\mathfrak{a}}^\dagger_t[F] = \check{\mathfrak{a}}_t[-\tau^1 F^*]  =  \check{\mathfrak{a}}_t[-\overline{F}] .
\end{equation}
In this sense, the creation operator for wavepacket $F$ is equivalent to the annihilation operator for the conjugate wavepacket $\overline{F}$, up to a minus sign.
Similarly, the equal-time commutation relations for two wavepacket operators are 
\begin{equation}
\label{eqn:mode-commutator}
    \left[ \check{\mathfrak{a}}_t[F], \check{\mathfrak{a}}_t^\dagger[G] \right] = \left( F, G \right).
\end{equation}
Thus, if $(F,F)>0$, $\check{\mathfrak{a}}_t[F]$ is a canonical annihilation operator and if $(F,F)<0$, it is a creation operator.

Evolving these operators in time may now be performed by employing the BdG kernel since 
\[
i\frac{d}{dt}\check{\mathfrak{a}}_t[F] = \left(F,i\partial_t \check{\Phi}(t) \right) = \left(F,\hat{\Omega}_{\tiny\textrm{BdG}} \check{\Phi}(t) \right).
\]
We now observe that with respect to this inner product, the BdG Hamiltonian obeys $\left(F, \hat{\Omega}_{\tiny\textrm{BdG}} G \right) = \left(\hat{\Omega}_{\tiny\textrm{BdG}} F,G \right)$, provided we respect the stationary-flow condition $\nabla\cdot(\rho\mathbf{v}) = 0$.
Using this property, we have 
\begin{equation}
\label{eqn:EOM-mode-operators}
i\frac{d\check{\mathfrak{a}}_t[F]}{dt} = \check{\mathfrak{a}}_t[\hat{\Omega}_{\tiny\textrm{BdG}} F].
\end{equation}
In particular, if we consider an energy eigenspinor $W_{\omega}$ with eigenvalue $\omega$, then we can solve Eqn.~\eqref{eqn:EOM-mode-operators} with    
\begin{equation}
\label{eqn:time-evolve}
\check{\mathfrak{a}}_t[W_{\omega}] = e^{-i\omega t} \check{\mathfrak{a}}[W_{\omega}] 
\end{equation}
from which we may obtain, e.g. the retarded and time-ordered correlation functions.

We may restrict ourselves to looking for only positive-frequency modes $W_{\omega\nu}$ (with $\nu$ indexing different degenerate modes) since the BdG Hamiltonian obeys the symmetry
\begin{equation}
\overline{\hat{\Omega}_{\tiny\textrm{BdG}}} \equiv \tau^1 \hat{\Omega}_{\tiny\textrm{BdG}}^*\tau^1 = - \hat{\Omega}_{\tiny\textrm{BdG}}.
\end{equation}
Thus up to a linear transformation amongst the degenerate eigenmodes, the mode with eigenvalue $-\omega$ is the conjugate of the mode with eigenvalue $+\omega$.
By judiciously choosing the basis elements $W_{\omega\nu}$ in each subspace, we can ensure that
\begin{equation}
\label{eqn:self-inversion}
    W_{-\omega\nu} =  - \overline{W}_{\omega\nu} \Leftrightarrow \check{\mathfrak{a}}_t[W_{-\omega\nu}] = \check{\mathfrak{a}}^\dagger_t[W_{\omega\nu}] .
\end{equation}

Rather than continue to analyze the problem at a general, abstract level, it will be beneficial to see how these ideas are applied to specific problems.
In particular, we will begin by considering the case of a homogeneous condensate, which may also be solved by means of the standard Bogoliubov transformation~\cite{AGD,Castin01}.
We will then move on to consider cases where the condensate possesses an event horizon.

\section{\label{sec:homgeneous}Homogeneous Condensate}
We begin by consider a translationally invariant stationary condensate. 
The BdG Hamiltonian is diagonalized in momentum space by the plane wave eigenmodes
\begin{equation}
\label{eqn:plane-wave}
W_{\mathbf{k}\sigma}(\mathbf{r}) = w_{\mathbf{k}\sigma}e^{i\mathbf{k}\cdot \mathbf{r}},
\end{equation}
where $\sigma$ indexes independent modes with the same momentum.
The BdG Hamiltonian is now a $2\times 2$ matrix in momentum space which reads
\begin{equation}
    \label{eqn:BdG-momentum}
    \Omega_{\tiny\textrm{{BdG}}}(\mathbf{k})= \tau^0\mathbf{v}\cdot\mathbf{k} + \tau^3 \left( \frac{\mathbf{k}^2}{2m} + g\rho \right) + i\tau^2 g\rho.
\end{equation}
At fixed energy $\omega>0$ we must solve the equation   
\begin{equation}
    \label{eqn:BdG-characteristic}
    \left[\tau^0\mathbf{v}\cdot\mathbf{k} + \tau^3 \left( \frac{\mathbf{k}^2}{2m} + g\rho \right) + i\tau^2 g\rho\right]w_{\mathbf{k}\sigma} = \omega w_{\mathbf{k}\sigma}
\end{equation}
for $\mathbf{k}$ and the corresponding spinor $w_{\mathbf{k}\sigma}$.
Setting the determinant to zero produces the well-known Bogoliubov dispersion relation
\begin{equation}
    \label{eqn:Bogoliubov-dispersion}
    \det\left[\Omega_{\tiny\textrm{{BdG}}}(\mathbf{k}) - \omega \right] = 0 \Rightarrow \left(\omega -\mathbf{k}\cdot\mathbf{v} \right)^2 = \frac{g\rho}{m}\mathbf{k}^2 + \left(\frac{\mathbf{k}^2}{2m}\right)^2, 
\end{equation}
from which we recognize the speed of sound $c^2 = \frac{g\rho}{m}$. 
In the long-wavelength limit, Eqn.~\eqref{eqn:Bogoliubov-dispersion} reduces to the Lorentz invariant dispersion $(\omega - \mathbf{v}\cdot\mathbf{k} )^2 - c^2 \mathbf{k}^2 \sim 0$.

\begin{figure}
    \centering
    \includegraphics[width=\linewidth]{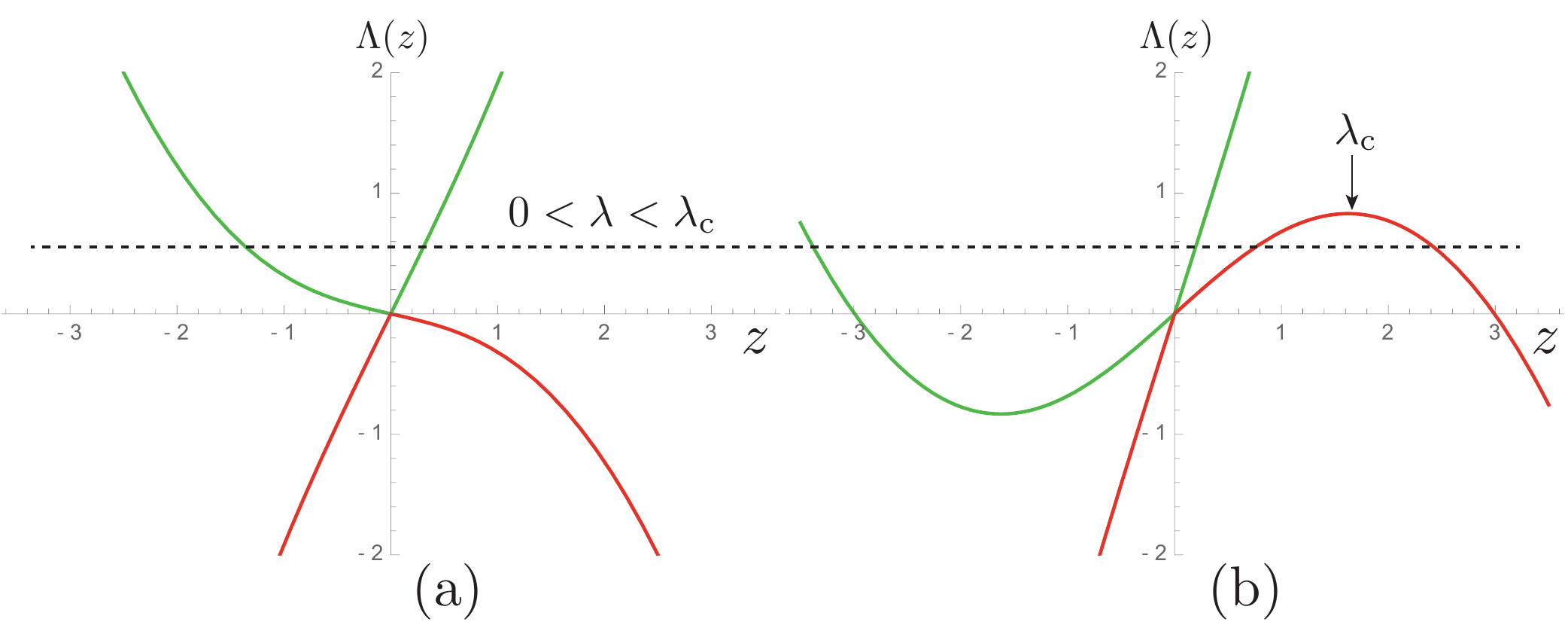}
    \caption{(Color Online) Dispersion relations solved graphically at fixed lab-frame energy (dashed line) for (a) subsonic case $\beta=.8$ and (b) supersonic case $\beta = 1.8$. 
    For $\beta^2 > 1$ there is a positive $\lambda_c$ such that within the window $0 < \lambda < \lambda_c$ there are four real solutions to the dispersion relation.
    For $\beta^2 >1$ but $\lambda > \lambda_c$, or when $\beta^2 <1$ there are only two real roots.}
    \label{fig:dispersions}
\end{figure}

We will henceforth restrict ourselves to the case of a one-dimensional system. 
In this case, Eqn.~\eqref{eqn:Bogoliubov-dispersion} becomes a quartic polynomial with roots $k_\nu(\omega)$, which must be real by normalizeability.
In solving this equation, it is convenient to introduce the unitless variables
\begin{equation} 
    \label{eqn:unitless}
    \begin{aligned} 
        &   z = \frac{k}{mc} \\
        &   \beta = \frac{v}{c}\\
        &   \lambda = \frac{\omega}{mc^2} \\
    \end{aligned} 
\end{equation} 
so that the eigenvalue problem now reads  
\begin{equation}
\label{eqn:BdG-unitless}
    \begin{aligned}
    &   \left[\beta z-\lambda + \tau^3 \left( 1+\frac12 z^2 \right) + i\tau^2 \right] w = 0     \\
    &   \left(1+\frac12 z^2 \right)^2-1 = \left(\lambda - \beta z\right)^2.   \\ 
    \end{aligned}
\end{equation}
The real roots can be found graphically by plotting the two functions  
\begin{equation}
\label{eqn:lab-frame-freq}
    \Lambda_\pm(z) = \beta z \pm \sqrt{z^2 + z^4/4}
\end{equation}
and finding their intersections with the prescribed lab-frame energy $\lambda$, as depicted in FIG.~\ref{fig:dispersions}.
In Eqn.~\eqref{eqn:lab-frame-freq} the $\pm$ sign determines the sign of the co-moving frequency $\lambda - \beta z$ and, as we will see later, the sign of the norm of the mode as defined in Eqn.~\eqref{eqn:norm}.

For $\beta^2 < 1$ the curve $\Lambda_\pm(z)$ is convex, as seen in FIG.~\ref{fig:dispersions}(a).
Thus there are only ever two real solutions, $z_{\pm p}$, both of which have positive co-moving frequency/norm but differ in their group velocity. 
The other two roots ($z_{\pm n}$) are complex conjugates and end up having negative norm (see Appendix~\ref{sec:norm-ev}).

In contrast, for $\beta^2 >1$ the curve $\Lambda_\pm(z)$ develops extrema at finite $\pm z_c$, found by solving the equation $\frac{\partial \Lambda_-(z) }{\partial z}\bigg|_{z_c} = 0$. 
For $\beta > 0$ this produces 
\begin{equation}
\label{eqn:xi-crit}
z_c = \sqrt{\frac{\beta^2}{2}\left( 1 + \sqrt{1+\frac{8}{\beta^2} } \right) - 2 } .
\end{equation}
As shown in FIG.~\ref{fig:dispersions}(b), for $0 < \lambda < \lambda_c(\beta) \equiv \Lambda_-(z_c)$ there will be four real roots to the dispersion relation. 
These additional roots are due to the effectively superluminal dispersion, which exhibits a group velocity that increases as momentum increases.
These two new roots $z_{\pm n}$ have negative co-moving frequency/norm and are again further labeled by the sign of their group velocity (we use positive for right-movers and negative for left-movers). 
Generically, one of the negative norm roots also linearly disperses at low energies (for $\beta > 0$ it is $z_{+n}$), identifying it as the phonon which travels upstream, though now it has been Doppler shifted to such a degree that its lab-frame energy and co-moving energy differ in sign.  

We now show that for real momenta the co-moving frequency indeed determines the norm of the mode. 
First, we obtain expressions for the eigenspinor
\begin{equation}
    \label{eqn:spinor-define}
    w_{\nu} = \left(\begin{array}{c}
    u_{\nu}   \\
    v_{\nu}   \\
    \end{array}\right)= \frac{1}{\sqrt{| 1 - |h_\nu |^2|}} \left(\begin{array}{c}
    1   \\
    h_\nu  \\
    \end{array}\right),
\end{equation}
with $h_{\nu}= \lambda - \beta z_\nu -(1+z_\nu^2/2)$.
We normalize this spinor to the inner product introduced in Eqn.~\eqref{eqn:norm}. 
When the momentum $z$ is real, this produces the result 
\begin{equation}
\label{eqn:spinor}
    w_{\nu} = \frac{1}{\sqrt{2|\lambda - \beta z_\nu|}} \left(\begin{array}{c}
    \left(-h_\nu\right)^{-\frac12} \\
    - \left(-h_\nu\right)^{+\frac12}\\
    \end{array}\right).
\end{equation}
Explicit calculation then confirms 
\begin{equation}
\label{eqn:norm-real}
    w_\nu^\dagger \tau^3 w_\nu = \textrm{sign}(\lambda - \beta z_\nu)
\end{equation}
which we recognize is simply the sign of the co-moving frequency, as claimed.

To conclude, we return to the position space eigenmodes and consider the inner product  
\[
(W_{\omega'\nu'}, W_{\omega\nu} ) = \int dx \rho e^{i\left( k_{\nu}(\omega) - k_{\nu'}(\omega')\right) x } w_{\omega'\nu'}^\dagger \tau^3 w_{\omega\nu},
\]
which evaluates to a delta function
\[
(W_{\omega'\nu'}, W_{\omega\nu} )  = 2\pi\rho \delta\left( k_{\nu}(\omega) - k_{\nu'}(\omega') \right).
\]
Note the appearance of the overall factor of the density. 
This implies the second-quantized operators for these momentum modes have the commutator 
\begin{equation}
    [ \check{\mathfrak{a}}_t[W_{\omega',\nu'}] , \check{\mathfrak{a}}^\dagger_t[W_{\omega,\nu}]] =  2\pi\rho \delta\left( k_{\nu}(\omega) - k_{\nu'}(\omega') \right).
\end{equation}
In order for this to match the canonical commutator we must divide by this factor of the density, so that the appropriately normalized position-space eigenmodes are in fact 
\begin{equation}
W_{\omega \nu}(x) = \frac{1}{\sqrt{\rho}} e^{i k_{\nu}(\omega) x} w_{\omega\nu}.
\end{equation}
This appearance of the factor of the density will prove to be important in the next section, where the density is spatially varying.

\section{\label{sec:step}Step-Like Horizon}

\subsection{Set Up}
Having examined the homogeneous system, we will now consider the ``simplest" generalization; a step-like discontinuity between two otherwise homogeneous regions. 
Specifically, the fluid profile considered is 
\begin{equation}
\label{eqn:step-profile}
\begin{aligned}
    & v(x) = \bigg\{\begin{aligned}
        & v_r   & x \geq 0\\
        & v_l   & x < 0 \\
        \end{aligned} \\
    & \rho(x) = \bigg\{\begin{aligned}
        & \rho_r   & x \geq 0\\
        & \rho_l   & x < 0 .\\
        \end{aligned} \\
\end{aligned}
\end{equation}
Though momentum is no longer a good quantum number, the lab-frame energy still is provided we maintain the stationary-flow condition.
In one dimension this requires 
\[
    \partial_x\left( \rho(x) v(x) \right) = 0 \Rightarrow \rho(x) v(x) = \textrm{constant.} 
\]
This constrains the step-profile from Eqn.~\eqref{eqn:step-profile} to obey
\begin{equation}
    \label{eqn:step-continuity}  
    \rho_l v_l = \rho_r v_r.
\end{equation}
It will be helpful to rewrite the local density $\rho(x)$ in terms of the local speed of sound $c(x) = \sqrt{g\rho(x)/m}$ which then implies that $c$ obeys
\begin{equation}
\label{eqn:continuity-sound}
    c_l^2 v_l = c_r^2 v_r.
\end{equation}
Thus, there are only three independent parameters amongst $v_l,c_l,v_r,c_r$.
We will parameterize these by the two independent unitless variables $\beta_l = v_l/c_l,\ \beta_r = v_r/c_r$ and $c_l$. 
This then fixes $c_r = \left(\frac{\beta_l}{\beta_r}\right)^{\frac13}c_l$. 

The one dimensional BdG Hamiltonian which governs the step system is  
\begin{equation}
    \label{eqn:BdG-step}
    \hat{\Omega}_{\tiny\textrm{BdG}} = \left( -\frac{1}{2m\rho(x)} \partial_x \rho(x)\partial_x +g\rho(x) \right)\tau^3 +g\rho(x) i\tau^2 -iv(x) \partial_x \tau^0 .
\end{equation}
Given the piecewise homogeneous nature of the Hamiltonian, we can solve for the eigenmodes of Eqn.~\eqref{eqn:BdG-step} by finding the appropriate plane-wave solutions in each half-space and then gluing them together at the interface, as is done for e.g. a particle reflecting off of a barrier.
The appropriate matching conditions may obtained by integrating Eqn.~\eqref{eqn:BdG-step} across the discontinuity (after multiplying by a factor of the density), and are 
\begin{equation}
    \label{eqn:matching}
    \begin{aligned}
        &   \left[W(x)\right]_{0^-}^{0^+} =  0  \\
        &   \left[\rho(x)\partial_x W \right]_{0^-}^{0^+}  = 0. \\ 
    \end{aligned}
\end{equation}
Each of these in turn produces two equations (recall that $W$ has two components) so that in total, Eqn.~\eqref{eqn:matching} presents four constraints.

We write the energy eigenmode as 
\begin{equation}
    \label{eqn:step-solution}
    W_{\nu}(x) = \sum_{\alpha }\Biggl\{ \begin{aligned}
    &   \frac{C^{\alpha l}_\nu}{\sqrt{\rho_l}} w_{\alpha}^{l}\exp\left(i k_{\alpha}^{l} x \right) & x < 0 \\
    &   \frac{C^{\alpha r}_\nu}{\sqrt{\rho_r}} w_{\alpha}^{r}\exp\left(i k_{\alpha}^{r} x \right) & x \geq 0 \\
    \end{aligned}
\end{equation}
where $\alpha$ now runs over all four solutions to the half-space homogeneous problem. 
Crucially, this includes the modes with complex momentum which have negative norm (see Appendix~\ref{sec:norm-ev}).
Within each half-space one of the complex negative-norm modes will describe an evanescent mode which is allowed by boundary conditions and must be included in order to solve the matching problem~\cite{Pavloff12,MacherParentani09D,Maksimov03,Rinaldi11}.
The other complex mode will describe an exponentially growing mode, which is forbidden (it will be formally convenient to include this mode but always set the coefficient to zero). 

The coefficients $C^{\alpha l/r}_\nu$ (whose dependence on $\omega$ has been suppressed for brevity) must now be chosen to satisfy the matching conditions Eqn.~\eqref{eqn:matching}.
We classify the eight $C$ coefficients by whether they are ingoing or outgoing.
For modes of real momentum, this is based on whether the lab-frame group velocity is directed towards or away from the step~\cite{Carusotto09}.
For modes of complex momentum, which don't have a group velocity, we instead treat a mode as outgoing if it is evanescent and ingoing if it is growing. 

If we hold $\beta_l < 1$ fixed then irrespective of $\omega > 0$, the $+pl,+nl$ modes are ingoing while the $-pl,-nl$ modes are outgoing.
Of these, the $+nl$ mode is growing, while the $-nl$ mode is evanescent.
As we vary $\beta_r$ on the other hand, we encounter two cases. 
The first case applies when either $0< \beta_r < 1$ or $\beta_r > 1$ but the frequency $\omega > \omega_c$, with 
\begin{equation}
    \label{eqn:step-cutoff}
    \omega_c = mc_r^2 \Lambda_-(z_c(\beta_r) )
\end{equation}
the cutoff frequency in the right half-plane.
In this case, the flow is effectively subsonic and the ingoing modes are $-pr,-nr$ while the $+pr,+nr$ modes are outgoing, with $+nr$ evanescent and $-nr$ growing.
This case is summarized in Table~\ref{tab:sub-sub-modes}.

The second possibility is that $\beta_r > 1$ and $0<\omega < \omega_c(\beta_l,\beta_r)$.
In this case the $\pm nr$ momenta become real and a new scattering channel opens. 
This regime is summarized in Table~\ref{tab:sub-sup-modes}.

\begin{table}[H]
    \centering
    \begin{tabular}{|c|c|c|c|}
    \hline
    Mode & Norm & Left Half-Space & Right Half-Space \\
    \hline
    $+p$  &   $+1$  &   \textrm{Right-mover (in)}    &   \textrm{Right-mover (out)} \\
    $-p$  &   $+1$  &   \textrm{Left-mover (out)}    &   \textrm{Left-mover (in)}   \\
    \hline
    $+n$  &   $-1$  &   \textrm{Growing (in)}    &   \textrm{Evanescent (out)}      \\
    $-n$  &   $-1$  &   \textrm{Evanescent (out)}    &   \textrm{Growing (in)}      \\
    \hline
    \end{tabular}
    \caption{Mode classification for step between two effectively subsonic regions at positive energy.}
    \label{tab:sub-sub-modes}
\end{table}
\begin{table}[H]
    \centering
    \begin{tabular}{|c|c|c|c|}
    \hline
     Mode & Norm & Left Half-Space & Right Half-Space \\
    \hline
    $+p$  &   $+1$  &   \textrm{Right-mover (in)}    &   \textrm{Right-mover (out)} \\
    $-p$  &   $+1$  &   \textrm{Left-mover (out)}    &   \textrm{Left-mover (in)}   \\
    \hline
    $+n$  &   $-1$  &   \textrm{Growing (in)}    &   \textrm{Right-mover (out)}      \\
    $-n$  &   $-1$  &   \textrm{Evanescent (out)}    &   \textrm{Left-mover (in)}      \\
    \hline
    \end{tabular}
    \caption{Mode classification for step between two subsonic regions at positive energy.
    Note that in the right-hand side, an ``ingoing" growing mode was converted into an ingoing scattering mode with real flux, and the corresponding outgoing evanescent mode was converted into an outgoing scattering mode with real flux.}
    \label{tab:sub-sup-modes}
\end{table}

\subsection{Solution}

We now apply the matching conditions in Eqn.~\eqref{eqn:matching}, which imposes four constraints. 
This linear system may be written as 
\begin{equation}
\label{eqn:matching-in-out}
\mathscr{M}_{\textrm{out}}\left(\begin{array}{c}
C^{+pr}  \\
C^{-pl}  \\
C^{+nr}  \\
C^{-nl}  \\
\end{array}\right)  =  \mathscr{M}_{\textrm{in}} \left(\begin{array}{c}
C^{+pl} \\
C^{-pr} \\
C^{+nl} \\
C^{-nr} \\
\end{array}\right),
\end{equation}
with the two matrices defined by 
\begin{equation}
\label{eqn:out-matrix}
    \mathscr{M}_{\textrm{out}}    =   
\left(\begin{array}{cccc}
    \beta^{\frac13}_r u_{+pr}           &  -\beta^{\frac13}_l u_{-pl}               &   \beta^{\frac13}_r u_{+nr}           &   - \beta^{\frac13}_l u_{-nl}             \\
    \beta^{\frac13}_r v_{+pr}           &  -\beta^{\frac13}_l v_{-pl}               &   \beta^{\frac13}_r v_{+nr}           &   - \beta^{\frac13}_l v_{-nl}             \\
    \beta^{-\frac23}_r z_{+pr} u_{+pr} &  -\beta^{-\frac23}_l z_{-pl} u_{-pl}     &   \beta^{-\frac23}_r z_{+nr} u_{+nr} &   - \beta^{-\frac23}_l z_{-nl} u_{-nl}   \\
    \beta^{-\frac23}_r z_{+pr} v_{+pr} &  -\beta^{-\frac23}_l z_{-pl} v_{-pl}     &   \beta^{-\frac23}_r z_{+nr} v_{+nr} &   - \beta^{-\frac23}_l z_{-nl} v_{-nl}   \\
\end{array}\right)
\end{equation}
\begin{equation}
\label{eqn:in-matrix}
    \mathscr{M}_{\textrm{in}}    =   
\left(\begin{array}{cccc}
    \beta^{\frac13}_l u_{+pl}           &  -\beta^{\frac13}_r u_{-pr}               &   \beta^{\frac13}_l u_{+nl}           &   - \beta^{\frac13}_r u_{-nr}             \\
    \beta^{\frac13}_l v_{+pl}           &  -\beta^{\frac13}_r v_{-pr}               &   \beta^{\frac13}_l v_{+nl}           &   - \beta^{\frac13}_r v_{-nr}             \\
    \beta^{-\frac23}_l z_{+pl} u_{+pl} &  -\beta^{-\frac23}_r z_{-pr} u_{-pr}     &   \beta^{-\frac23}_l z_{+nl} u_{+nl} &   - \beta^{-\frac23}_r z_{-nr} u_{-nr}   \\
    \beta^{-\frac23}_l z_{+pl} v_{+pl} &  -\beta^{-\frac23}_r z_{-pr} v_{-pr}     &   \beta^{-\frac23}_l z_{+nl} v_{+nl} &   - \beta^{-\frac23}_r z_{-nr} v_{-nr}   \\
\end{array}\right).
\end{equation}
Note we have used the unitless variables introduced in Eqn.~\eqref{eqn:unitless}, now given on each half-space (though the continuity relation constrains them, in general). 
Normalizeability requires the coefficients of the ingoing negative norm modes ($-nr,+nl$) be set to zero if their momentum is complex. 
This is always the case for the $+nl$ mode, but for the $-nr$ mode this depends on whether $\omega$ is less than the cutoff $\omega_c$ or not.

Inverting the $\mathscr{M}_{\textrm{out}}$ matrix produces
\begin{equation}
\label{eqn:matching-S-define}
\left(\begin{array}{c}
C^{+pr}  \\
C^{-pl}  \\
C^{+nr}  \\
C^{-nl}  \\
\end{array}\right)  =  \mathcal{A} \left(\begin{array}{c}
C^{+pl} \\
C^{-pr} \\
C^{+nl} \\
C^{-nr} \\
\end{array}\right).
\end{equation}
with the $\mathcal{A}$ matrix defined by 
\begin{equation}
    \label{eqn:S-matrix}
    \mathcal{A} \equiv \mathscr{M}_{\textrm{out}}^{-1} \mathscr{M}_{\textrm{in}},
\end{equation}
which determines the amplitudes of the various outgoing modes present in a particular energy eigenmode, given the initial ingoing amplitudes. 

When the step is between two subsonic flows, both the $C^{+nl}$ and $C^{-nr}$ coefficients must be set to zero.
Thus, there are only two degenerate eigenmodes which correspond to a modes incident from the left and right.
In this sense, the subsonic-subsonic step may be considered as being ``adiabatically" connected to the homogeneous system, where the matrix $\mathcal{A}$ becomes a trivial identity map. 

For a step between a subsonic flow and a supersonic flow, when the energy is below the cutoff $\omega_c$ an event horizon appears and the $\pm nr$ modes become scattering states.
These new scattering channels produce a third degenerate eigenmode, increasing the rank of the scattering matrix at this energy from two to three.
Because this mode converts an incident negative norm wave into an outgoing positive norm component, it is responsible for producing Hawking radiation~\cite{Pavloff12,Rinaldi11}. 
It would interesting to determine under what general circumstances the rank of the scattering matrix may be related to the presence of event horizons in the spacetime.
In addition, the nature of the transition from rank two to rank three may be interesting to study, as it seems impossible for it to occur in a smooth manner.
We will leave these questions open for future studies.

\begin{figure}
    \centering
    \includegraphics[width=\linewidth]{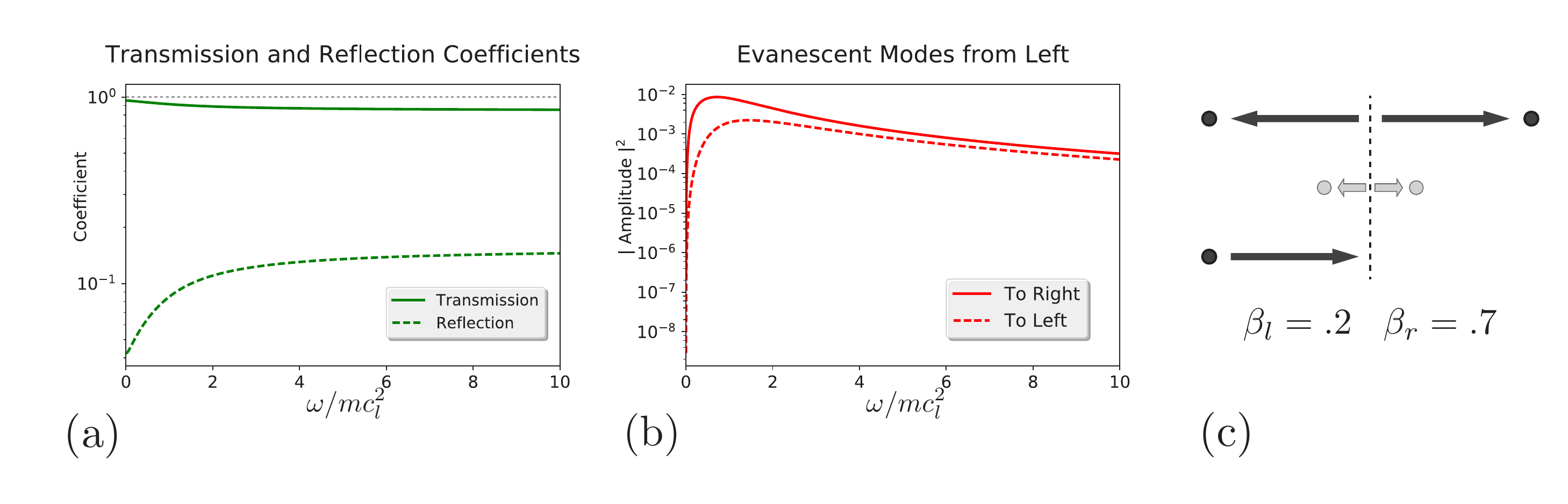}
    \caption{(Color Online) (a) Scattering coefficients as functions of energy for a step between two subsonic flows, and the corresponding evanescent mode amplitudes (b).
    A schematic describing the process (c), with dark arrows representing positive norm modes and light gray arrows representing evanescent modes.
    The direction of the arrowhead signifies whether the mode is considered ingoing or outgoing.}
    \label{fig:sub-sub-scatter}
\end{figure}

Though not our main focus, for completeness we will now explicitly obtain the scattering ($\mathcal{S}$) matrix.
This matrix is obtained from the matrix $\mathcal{A}$ by weighting each mode by its asymptotic conserved current, as per equation~\eqref{eqn:operator-current}.
Since the current is defined as the value at spatial infinity, evanescent modes do not carry a well-defined flux, nor do they enter into the unitarity expression.
For a scattering mode, the asymptotic current it carries is 
\begin{equation}
    \label{eqn:mode-current}
    J^{\alpha} = w_\alpha^\dagger \left( v \tau^3 + \frac{k_\alpha}{m} \tau^0\right) w_{\alpha}.
\end{equation}
It may be shown (see Appendix~\ref{sec:current-group}) that this current is equal to the group velocity of the mode, weighted by its norm.
Thus, the direction of current flow may be determined graphically as well.
For the subsonic-subsonic configuration, unitarity requires 
\begin{equation}
    \label{eqn:sub-sub-unitarity}
    \begin{aligned}
    &   \bigg|\frac{J^{+pr}}{J^{+pl}}\bigg||\mathcal{A}^{+pr}_{+pl}|^2 + \bigg|\frac{J^{-pl}}{J^{+pl}}\bigg||\mathcal{A}^{-pl}_{+pl}|^2 = 1 \\
    &   \bigg|\frac{J^{+pr}}{J^{-pr}}\bigg||\mathcal{A}^{+pr}_{-pr}|^2 + \bigg|\frac{J^{-pl}}{J^{-pr}}\bigg||\mathcal{A}^{-pl}_{-pr}|^2 = 1. \\
    \end{aligned}
\end{equation}
In FIG.~\ref{fig:sub-sub-scatter}, the reflection and transmission coefficients are plotted as functions of the lab-frame energy for a mode incident from the left (exterior).

For the subsonic-supersonic configuration (below threshold) we take into account the additional negative norm scattering modes, producing the unitarity relations 
\begin{equation}
    \label{eqn:sub-sup-unitarity}
    \begin{aligned}
    &   \bigg|\frac{J^{+pr}}{J^{+pl}}\bigg||\mathcal{A}^{+pr}_{+pl}|^2 + \bigg|\frac{J^{-pl}}{J^{+pl}}\bigg||\mathcal{A}^{-pl}_{+pl}|^2 - \bigg|\frac{J^{+nr}}{J^{+pl}}\bigg||\mathcal{A}^{+nr}_{+pl}|^2= 1 \\
    &   \bigg|\frac{J^{+pr}}{J^{-pr}}\bigg||\mathcal{A}^{+pr}_{-pr}|^2 + \bigg|\frac{J^{-pl}}{J^{-pr}}\bigg||\mathcal{A}^{-pl}_{-pr}|^2 - \bigg|\frac{J^{+nr}}{J^{-pr}}\bigg||\mathcal{A}^{+nr}_{-pr}|^2= 1 \\
    &   \bigg|\frac{J^{+pr}}{J^{-nr}}\bigg||\mathcal{A}^{+pr}_{-nr}|^2 + \bigg|\frac{J^{-pl}}{J^{-nr}}\bigg||\mathcal{A}^{-pl}_{-nr}|^2 - \bigg|\frac{J^{+nr}}{J^{-nr}}\bigg||\mathcal{A}^{+nr}_{-nr}|^2= -1 \\
    \end{aligned}.
\end{equation}
Similar relations have been obtained in, e.g.~\cite{Pavloff12,Pavloff13,Carusotto09,Parentani10}.
Note that the third scattering channel has an overall minus sign, due to the incoming mode having overall negative norm.
The presence of the outgoing negative norm states implies that the reflection and transmission coefficients together sum to a value larger than unity, a hallmark of supperradiance.

\begin{figure}
    \centering
    \includegraphics[width=\linewidth]{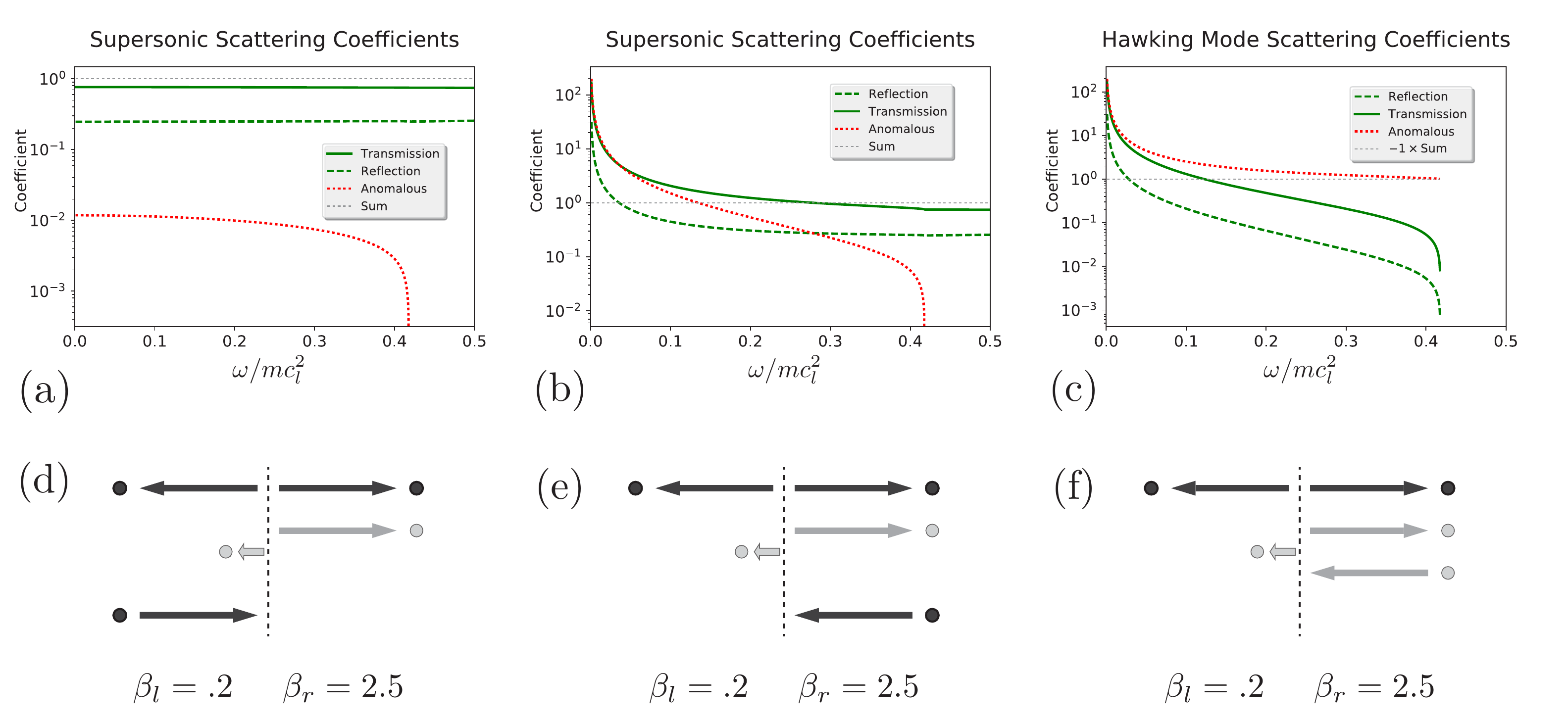}
    \caption{(Color Online) Scattering coefficients for subsonic-supersonic step as a function of lab-frame energy.
    (a) Scattering of a particle incident from the left.
    The anomalous mode corresponds to conversion into an outgoing negative norm mode. 
    Above the threshold energy, this coefficient goes to zero, as the outgoing channel becomes evanescent.
    (b) Scattering of a particle incident from the right, a process which can occur due to the superluminal dispersion.
    (c) The Hawking mode, whereby an incident negative norm mode scatters off of the event horizon.
    This ingoing channel becomes an exponentially growing mode above the threshold energy, where all the coefficients go to zero.
    In the schematics (d-f), the large grey arrows indicate the negative norm scattering states.}
    \label{fig:sub-sup-scatter}
\end{figure}

In FIG.~\ref{fig:sub-sup-scatter} we depict the scattering coefficients for each ingoing configuration as a function of energy.
Below the cutoff energy there are three ingoing configurations, each scattering into the three possible outgoing channels.
Of these three ingoing channels, two have an overall positive norm, while the third describes the Hawking channel and an has overall negative norm.
The Hawking radiation spectrum is determined by the transmission coefficient which describes the scattering of this mode into the outgoing positive norm mode outside the horizon (the $-pl$ mode). 

\begin{figure}
    \centering
    \includegraphics[width=\linewidth]{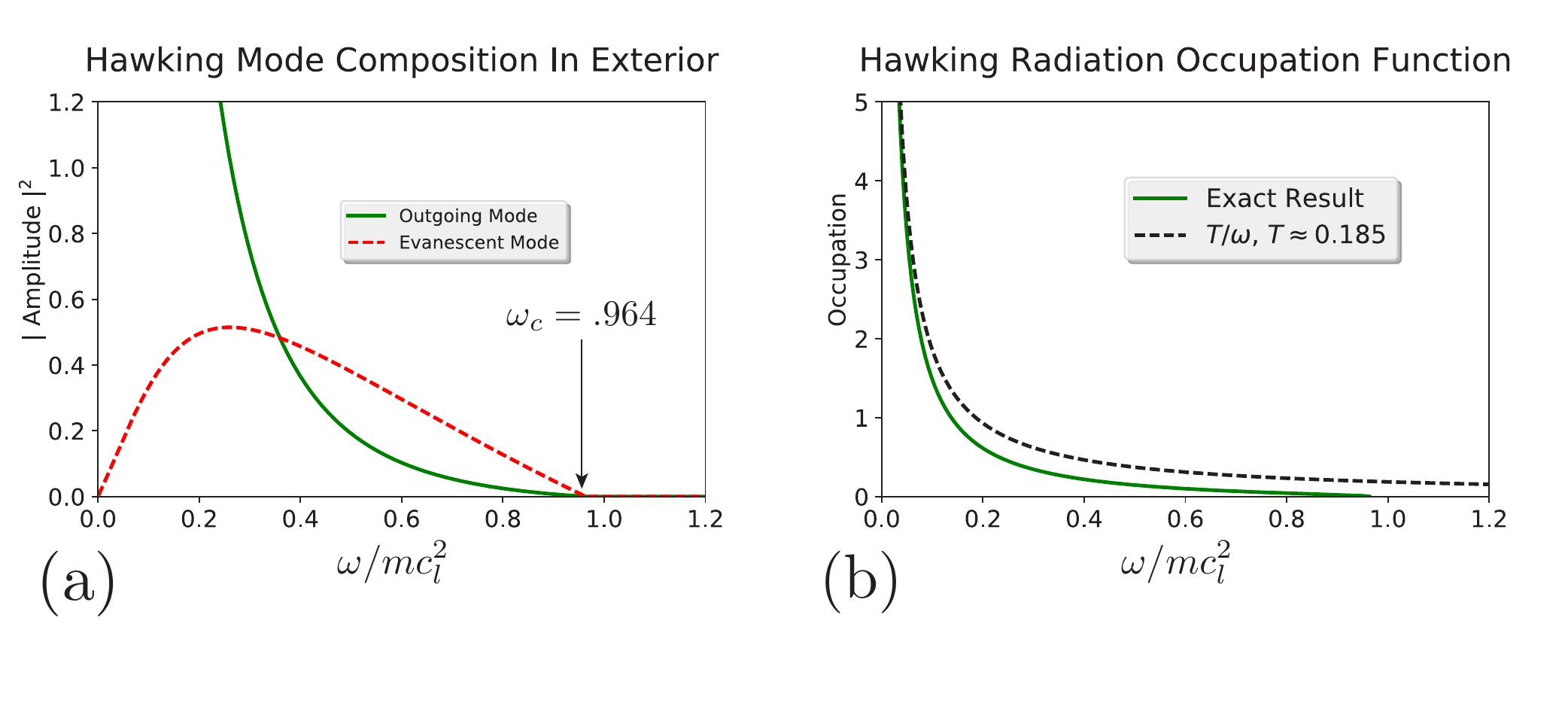}
    \caption{(Color Online) (a) The squared-amplitudes of the various modes comprising the Hawking eigenmode for $x<0$ ($\beta_l = .7,\beta_r = 2.5$). 
    We see that they both go to zero at the cutoff energy, which is also depicted.
    (b) The computed Hawking flux compared to an approximation by $\sim 1/\omega$, the classical (low energy) approximation to thermal occupation. 
    The proportionality coefficient is extracted and represents (up to an overall greybody absorption factor) the effective black hole temperature.
    While the two curves agree at low frequencies, they clearly disagree at higher energies.}
    \label{fig:hawking-flux}
\end{figure}

This Hawking flux exiting the black hole is depicted in more detail in FIG.~\ref{fig:hawking-flux}, where it is also compared to the magnitude of the evanescent mode present in this eigenmode. 
While the evanescent mode amplitude vanishes at zero energy, the Hawking flux diverges as $1/\omega$ at low frequencies, reflecting its effectively thermal distribution at low energies.
As claimed earlier, the distribution function departs from thermality at higher energies before vanishing at the cutoff energy $\omega_c$.
We now shift our focus to the evanescent mode, which only exists close to the event horizon. 

\section{\label{sec:evanescent}Evanescent Modes}

In the previous section we argued that evanescent modes do not contribute to the scattering relations since they carry no asymptotic flux.
One may then wonder under what conditions they are physically important.
We will now demonstrate that if one considers observables which depend on the near-horizon correlation functions~\cite{Carusotto08,Kravtsov09,Rinaldi11}, the evanescent modes will be important as they will modify quantities measured near the horizon.
As a simple example, we consider the norm density of a particular eigenmode, a quantity which is related to the quantum density fluctuations.

\begin{figure}
    \centering
    \includegraphics[width=\linewidth]{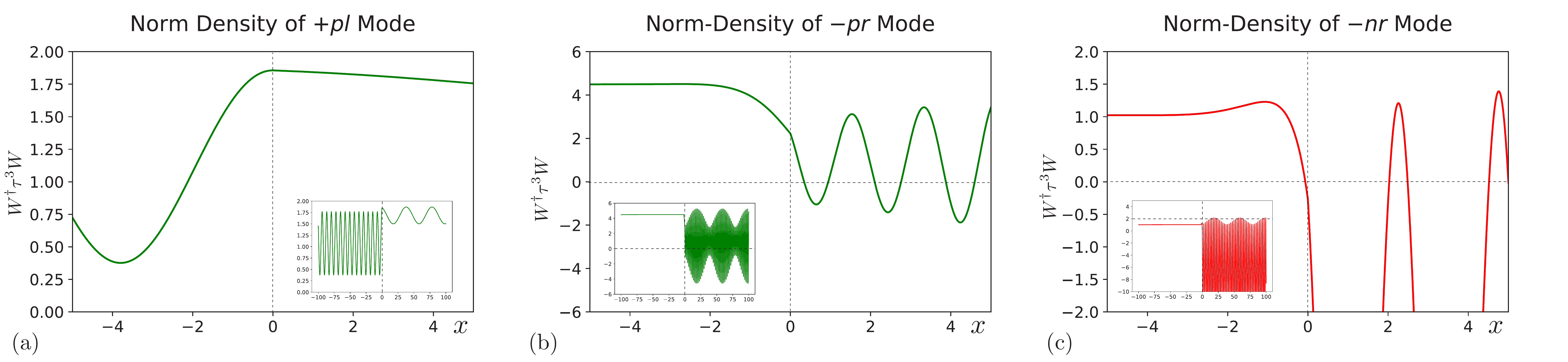}
    \caption{(Color Online) Plot of $W^\dagger(x) \tau^3 W(x)$ as a function of $x$ for a particular eigenvalue ($\omega = .260, \beta_l = .7,\beta_r = 2.5$). 
    (a) The near horizon density of the $+pl$ mode, which is incident from the left. 
    The inset of (a) shows the same, but plotted over a larger range of $x$.
    (b) The near horizon density of the $-pr$ mode, which is a superluminal positive norm mode incident from right. 
    The inset shows a larger range of $x$. 
    The evanescent mode is seen as the modulation of the density at $x\lesssim 0$ away from its asymptotic value.
    (c) The near horizon density of the $-nr$ mode, which is the Hawking mode with a negative norm mode incident from the right. 
    The inset shows a larger range of $x$. 
    The evanescent mode is again noticeable here as a modulation of the density for $x\lesssim 0$. 
    The horizontal and vertical dashed lines indicate zero for $W^\dagger\tau^3 W$ and $x$, where the horizon is located, respectively.
    It should be noted that the true conserved density is $\rho W^\dagger \tau^3 W$, rather than $W^\dagger \tau^3 W$ which has been plotted instead for clarity. 
    Statements made about the evanescent mode are unaffected by this distinction.}
    \label{fig:densities}
\end{figure}

Specifically, we will study the Hawking ($-nr$) mode, which only has one scattering mode outside of the horizon.
As such, far from the horizon the norm density is a featureless constant, with value $\frac{1}{c_l^2}|\mathcal{A}^{-pl}_{-nr}|^2$.
However, near the horizon the evanescent mode is non-zero and can exhibit quantum interference with the outgoing Hawking flux, leading to a deviation of the norm density from its value inferred by an observer at spatial infinity.
This interference is clearly visible in FIG.~\ref{fig:densities}(b),(c) where the density for $x\rightarrow -\infty$ is constant and featureless, while the density near $x \lesssim 0$ deviates from this value quite significantly just outside the event horizon.

When considering near-horizon physics, the evanescent modes will have important contributions which would otherwise be overlooked if the horizon curvature is small, or if the observer is sufficiently far from the horizon. 
Such behavior is not permissible by the equivalence principle, which implies that observers crossing the event horizon experience a locally flat spacetime.
It is not surprising therefore that these modes decay over a length scale governed by the scale at which Lorentz invariance is violated. 
Using the Bogoliubov dispersion relation we find that the imaginary part of the evanescent mode momentum is related to the real part by  
\begin{equation}
    \label{eqn:imag-part}
    \left(\textrm{Im}(z_{-nl})\right)^2 = \left(\textrm{Re}(z_{-nl}) \right)^2 + 2(1-\beta_l^2) + \frac{2\beta_l \lambda_l }{\textrm{Re}(z_{-nl})}.
\end{equation}
The first term and third terms can be shown to be non-negative for $\beta,\lambda>0$.
Thus, we can bound the imaginary part from below by 
\begin{equation}
    \label{eqn:imag-bound}
    \textrm{Im}(z_{-nl}) \geq \sqrt{2(1-\beta_l^2)} .
\end{equation}
Replacing units, we find that the evanescent modes are forced to decay over a length scale $L_{-nl}$ such that
\begin{equation}
    \label{eqn:decay-bound}
    L_{-nl} \leq \frac{1}{\sqrt{2}mc_l\sqrt{(1-\beta_l^2)} }.
\end{equation}
For any finite subsonic flow this scale is finite and furthermore, as $m \rightarrow \infty$ (at fixed $c_l$), this length scale goes to zero, completely removing the evanescent modes from the spectrum.
This is inline with our intuition since the mass $m$ effetively sets the scale at which the superluminal dispersion ruins Lorentz invariance; thus, this limit corresponds to enforcing Lorentz symmetry throughout the entire spectrum.
In this case, the equivalence principle requires the evanescent modes to disappear, as they do.

\begin{figure}[H]
    \centering
    \includegraphics[width=\linewidth]{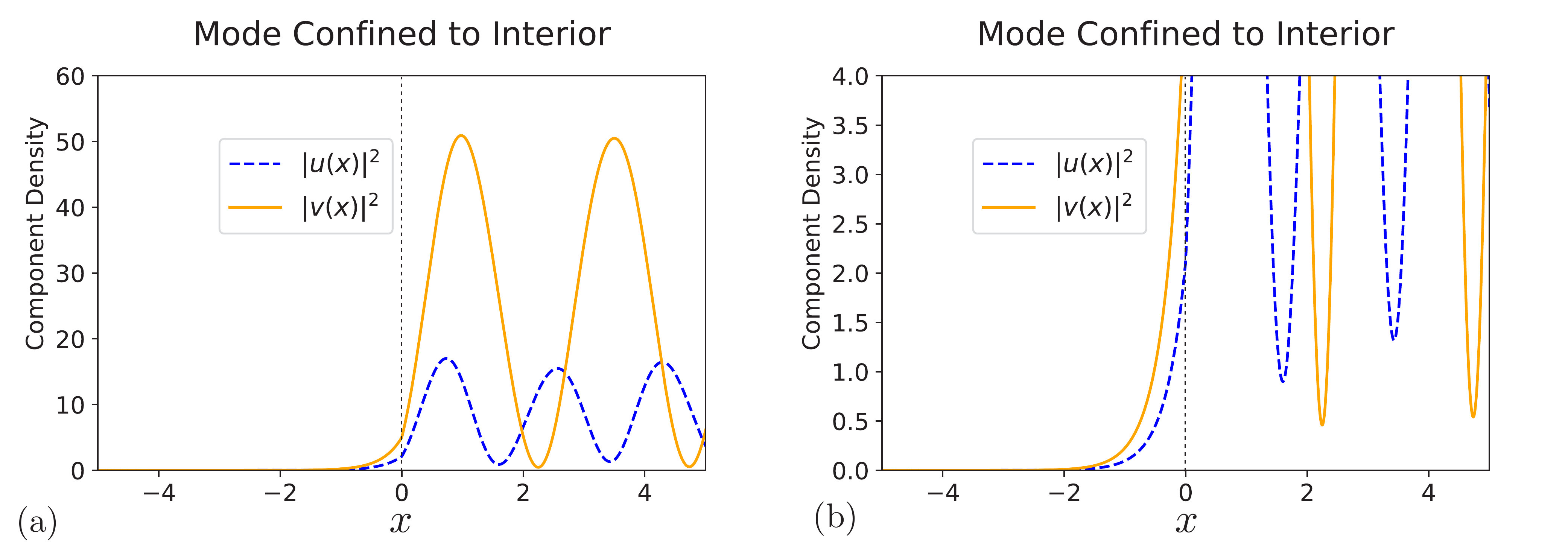}
    \caption{(Color Online) Confined mode for horizon with $\beta_l = .7, \beta_r = 2.5$ for a particular eigenenergy ($\omega = .260$).
    (a) The density (modulus squared) of each component of the mode confined to the interior of the horizon. 
    Note that both components independently decay to zero for $x <0$ (outside the event horizon).
    (b) The same quantity, plotted over a smaller y-scale, emphasizing the scale of the decay for $x<0$. }
    \label{fig:confined-mode}
\end{figure}

To emphasize the potential importance of these evanescent modes, we recall that in the presence of an event horizon the energy eigenbasis becomes three-dimensional.
In this case, we can form the linear combination of ingoing-eigenmodes 
\begin{equation}
    \label{eqn:confined-mode}
   W_{\textrm{confined}}(x) \equiv  \frac{\mathcal{A}_{-pr}^{-pl}W_{-nr}(x) -  \mathcal{A}_{-nr}^{-pl}W_{-pr}(x)}{\mathcal{A}_{-pr}^{-pl}\mathcal{A}_{-nr}^{-nl} - \mathcal{A}_{-nr}^{-pl}\mathcal{A}_{-pr}^{-nl}}, 
\end{equation}
which describes a coherent superposition of the Hawking mode and an incident superluminal particle.
This particular combination of modes has no flux escaping the black hole as the two incident amplitudes coherently cancel outside the event horizon.
Nevertheless, due to the evanescent modes it is seen to have finite support outside of the event horizon.
This is illustrated in FIG.~\ref{fig:confined-mode}, which depicts the norm density and individual components of a particular confined eigenmode, indeed confirming that it is exponentially localized to the interior of the black hole.
It appears that if we were to extend the black hole interior to include a white hole, this confined mode would represent ``half" of a black hole bound state, which are the modes responsible for black-hole-lasing and dynamical instability~\cite{Jacobson99,Parentani10,Leonhardt18,Steinhauer14}. 
Analyzing the stability and evolution of this confined mode with and without the accompanying white hole may uncover interesting instabilities which develop conditioned on the nature of the black hole interior.

\section{\label{sec:conclusion}Conclusion}
In this work we have systematically studied the step-like horizon formed in a quasi-one dimensional flowing BEC, which we argue serves as a model for acoustic black holes of very large curvature.
In addition to computing the scattering coefficients of this system (including the Hawking flux coefficient), we have also highlighted and studied the properties of the evanescent modes which form at the event horizon and result from model's non-linear dispersion.
These evanescent modes have been conclusively shown to modify near-horizon observables, despite the fact that they do not affect the scattering relations.

Given that the evanescent modes are effectively negative norm states tunneling across the event horizon, and have no flux out to infinity, it is interesting to speculate on what role, if any, these modes may serve in resolving the black hole information paradox. 
Through their effect on observables (such as the density and current fluctuations) near the horizon, it is conceivable that they may provide a route for information trapped behind the horizon to escape.
In particular, though they cannot asymptotically carry any information away as they are only a virtual process, it may be possible to retrieve information from the interior via an ``external measurement" of the system (e.g. a projective measurement of the density).
This is especially important since, in our model we have neglected interactions between the Bogoliubov quasiparticles, truncating the Heisenberg equations to linear order in fluctuations.
In the presence of quasiparticle interactions, it is plausible that the system will ``self-measure," as the virtual evanescent modes collide and interact with the outgoing Hawking flux, leading to a genuine leakage of information out of the event horizon.

In a similar vein, it is interesting to note that even though the condenate varies in an abrupt step-like manner, the evanescent modes seem to modulate observables on a longer length scale which is comparable to their decay length.
If we assume that the quasiparticle correction to the physical boson density behaves similarly, then we should expect that an initially sharp step-like condensate will become dressed by the evanescent modes, smearing it into a less abrupt horizon.
This is interesting since it potentially provides an example where the quasiparticle back-reaction on the condensate modifies the event horizon itself.
When quasiparticle interactions are included this may generate entanglement between the event horizon and the outgoing Hawking radiation. 

Finally, we note that evanescent modes have been seen in other contexts, e.g.the AdS-CFT correspondence~\cite{Wintergerst}, where evanescent modes are also seen to emanate from an apparent AdS black hole event horizon.
Given that the model we consider only has emergent Lorentz invariance, it is interesting to consider whether it is possible to connect the analogue of the AdS-CFT correspondence for a theory with only emergent Lorentz invariance, i.e. one with a sonic black hole in its bulk and some form of approximate-CFT on the boundary.
Similarly, whether it is possible to see these evanescent features in models of quantum gravity with emergent Lorentz invariance (e.g. Ho\v{r}ava gravity~\cite{Horava09,Wang17} ) is another interesting avenue of research. 

\begin{acknowledgments}
The authors would like to acknowledge productive discussions with Justin Wilson, Aydin Keser, Greg Gabadadze, Brian Swingle, and Iacopo Carusotto.
This work was supported by the National Science Foundation Graduate Research Fellowship Program under Grant No. DGE 1322106 (JC) and US-ARO (contract No. W911NF1310172), NSF-DMR 1613029, and Simons Foundation (VG). 
GR is grateful for the IQIM, an NSF Frontier center, and both GR and VG are grateful for the hospitality of the Aspen Center for Physics, which is supported by National Science Foundation grant PHY-1607761, where part of the work was done.
\end{acknowledgments}

\appendix
\section{\label{sec:norm-ev}Norm of Complex Modes}
\begin{figure}[H]
    \centering
    \includegraphics[width=.75\linewidth]{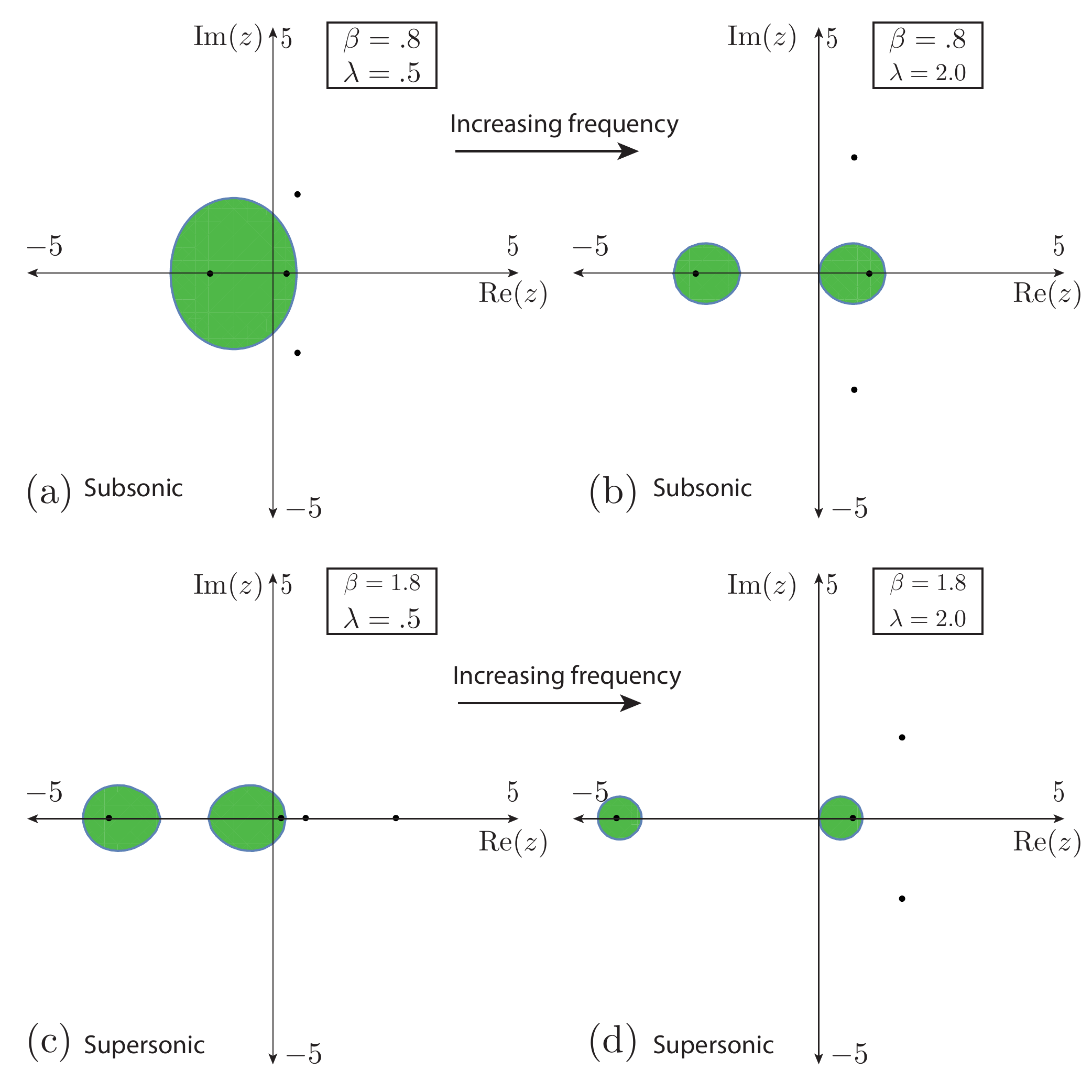}
    \caption{(Color Online) Regions in the complex $z$ (momentum) plane where the spinor has positive norm are colored green.
    Figures (a), (b) are for a subsonic flow at low and high frequencies, respectively. 
    Figures (c), (d) are for a supersonic flow at low and high frequencies.
    Each of the four roots for the given $\beta,\lambda$ are shown in the complex plane. 
    Two are real and always fall in the green region (the $\pm p$ roots) while two always fall outside (the $\pm n$ roots).
    When the $\pm n$ momenta are not real, this shows that they still have negative norm.
    }
    \label{fig:complex-roots}
\end{figure}

Here we demonstrate that for the homogeneous system the complex momentum modes have negative norm.
We begin with the unitless BdG equation
\begin{equation}
    \left[ (1+\frac12 z^2)\tau^3 + i\tau^2 + \beta z - \lambda\right] w = 0.
\end{equation}
There are four roots to the characteristic equation, $z_{\pm p}$ which are the two positive norm roots, and $z_{\pm n}$ which are either negative norm and real or complex conjugate pairs. 
The corresponding spinors are 
\begin{equation}
    w_\nu = \frac{1}{\sqrt{|1-|h_\nu|^2|}}\left(\begin{array}{c}
    1   \\
    h_\nu\\
    \end{array}\right)
\end{equation}
where $h_\nu = \lambda - \beta z - (1+\frac12 z^2)$, as is the case for the real-momentum modes.
These modes have negative norm whenever 
\begin{equation}
    |h_\nu|^2 > 1 \Rightarrow | \lambda - \beta z_\nu - (1+\frac12 z_\nu^2 ) |^2 > 1.
\end{equation} 
If we relax the constraint that $z = z_\nu$, one of the roots, we can study the regions defined by this inequality in the complex $z$ plane, for fixed values of $\beta>0, \lambda >0$.
We can then determine, for this $\beta,\lambda$, what $z_\nu$ is and see which region of the complex plane it falls in.
This is depicted in FIG.~\ref{fig:complex-roots}, for a number of different parameter values.

\section{\label{sec:current-group}Current and Group Velocity}
Here we show that for a scattering mode the group velocity and current are equivalent.
We first consider a momentum eigenmode,
\[
W_{\mathbf{k}}(\mathbf{r}) = e^{i\mathbf{k}\cdot\mathbf{r}} \frac{w_{\mathbf{k}}}{\sqrt{\rho}}
\]
which obeys the momentum space BdG equation 
\[
\left( \tau^3 \left( \frac{\mathbf{k}^2}{2m} + mc^2 \right) + i\tau^2 mc^2 + \mathbf{v}\cdot\mathbf{k} - \omega \right)w_{\mathbf{k}} = 0.
\]
We differentiate with respect to the wave-vector to get 
\begin{equation}
\left( \tau^3 \frac{\mathbf{k}}{m}+ \mathbf{v} - \frac{\partial \omega}{\partial \mathbf{k}} \right)w_{\mathbf{k}}  +\left( \tau^3 \left( \frac{\mathbf{k}^2}{2m} + mc^2 \right) + i\tau^2 mc^2 + \mathbf{v}\cdot\mathbf{k} - \omega \right)\frac{\partial w_{\mathbf{k}}}{\partial \mathbf{k}} = 0.
\end{equation}
We now apply $w_{\mathbf{k}}^\dagger \tau^3 $ from the left and use the Hermiticity of the BdG Hamiltonian with respect to the $\tau^3$ inner-product to eliminate the term involving $\frac{\partial w}{\partial \mathbf{k}}$.
This then produces the result 
\begin{equation}
    w_{\mathbf{k}}^\dagger \tau^3 w_{\mathbf{k}} \frac{\partial \omega}{\partial \mathbf{k}} = w_{\mathbf{k}}^\dagger \left[ \mathbf{v}\tau^3 + \frac{\mathbf{k}}{m}\right]w_{\mathbf{k}},
\end{equation}
which is the desired relation between group-velocity (LHS), and norm current (RHS).
This also will potentially generalize the concept of group velocity to the evanescent modes, which still have a well-defined norm current.

\bibliography{references}

\end{document}